\documentclass[twocolumn,prb,floatfix,showpacs,superscriptaddress,amsmath,amssymb,letterpaper]{revtex4}
\usepackage{graphicx}
\usepackage{bm}

\begin{document}
\title{Magnetic properties of the quasi-two-dimensional $S=\frac{1}{2}$ Heisenberg antiferromagnet [Cu(pyz)$_2$(HF$_2$)]PF$_6$}
\author{E. \v{C}i\v{z}m\'{a}r}
\affiliation{Hochfeld-Magnetlabor Dresden (HLD), FZ Dresden-Rossendorf,
D-01314 Dresden, Germany}
\affiliation{Centre of Low Temperature Physics, P.J. \v{S}af\'{a}rik
University, SK-041 54 Ko\v{s}ice, Slovakia}
\author{S. A. Zvyagin}
\author{R. Beyer}
\author{M. Uhlarz}
\author{M. Ozerov}
\author{Y. Skourski}
\affiliation{Hochfeld-Magnetlabor Dresden (HLD), FZ Dresden-Rossendorf,
D-01314 Dresden, Germany}
\author{J. L. Manson}
\affiliation{Department of Chemistry and Biochemistry, Eastern
Washington University, Cheney, WA 99004, USA}
\author{J. A. Schlueter}
\affiliation{Materials Science Division, Argonne National Laboratory,
Argonne, IL 60439, USA}
\author{J. Wosnitza}
\affiliation{Hochfeld-Magnetlabor Dresden (HLD), FZ Dresden-Rossendorf,
D-01314 Dresden, Germany}
\date{\today}

\begin{abstract}

We report on high-field magnetization, specific-heat and electron spin
resonance (ESR) studies of the quasi-two-dimensional spin-1/2 Heisenberg
antiferromagnet [Cu(pyz)$_2$(HF$_2$)]PF$_6$. The frequency-field diagram  of ESR modes below $T_N= 4.38$~K is described in the frame of the mean-field theory, confirming a~collinear magnetic structure with an easy-plane anisotropy. The obtained results allowed us to determine the anisotropy/exchange interaction ratio, $A/J = 0.003$, and the upper limit for the inter/intra-plane exchange-interaction ratio, $J'/J = 1/16$.
It is argued that despite the onset of 3D long-range magnetic ordering the magnetic properties of this material (including high-magnetic-field magnetization and non-monotonic field dependence of the N\'{e}el temperature) are strongly affected by two-dimensional spin correlations.
\end{abstract}

\maketitle

\section{Introduction}
Recently, a~considerable amount of attention has been
devoted to the theoretical and experimental investigation of two-dimensional
(2D) quantum spin systems. In case of the ideal 2D Heisenberg antiferromagnet
(AF) on a~square lattice, the magnetic long-range order is suppressed by zero-point fluctuations
at any finite temperature.\cite{mw} However, the presence of an easy-axis
anisotropy can induce a~finite-temperature phase transition into the N\'{e}el
ordered state.\cite{cuccoli1} On the other hand, for an easy-plane anisotropy, a~transition of the Berezinski-Kosterlitz-Thouless (BKT) type~\cite{cuccoli1,berez,kost} was proposed, where the
high-temperature disordered phase can be described as a~gas of vortices.
Below the BKT transition ($T_{BKT}$), the vortices
are bound in vortex-antivortex pairs, and the spin-spin correlation
decay changes from exponential to algebraic.\cite{Gupta} A~crossover from the isotropic to XY behavior in low fields has been predicted for 2D Heisenberg AF.\cite{pires, cuccoli2}

Interplane interactions (which are always present in real materials)
can significantly modify the ground-state properties of quasi-2D
systems, inducing for instance a~phase transition with three-dimensional
(3D) long-range magnetic order. Due to the onset of 3D ordering in
most real materials, the critical behavior of diverging quantities as
anticipated for ideal 2D systems at $T_{BKT}$ is hardly observable
experimentally.

The quasi-2D Heisenberg AF model with interplane interactions can be
described by the Hamiltonian
\begin{equation}
   \mathcal{H}=J {\sum_{<i,j>_{a\space b}} {\bf{S}_i \cdot \bf{S}_j}+J'
   \sum_{<i,j>_{c}} {\bf{S}_i \cdot \bf{S}_j}}-\beta\sum_i {S_i^z},
   \label{eq:Hamiltonian}
\end{equation}
with $\beta =g\mu_{\text{B}}B$, where $\mu_B$ is the Bohr magneton,
$J$ and $J'$ are the intra- and interplane exchange interactions, respectively,
and ${<i,j>_{a\space b}}$ and ${<i,j>_{c}}$ correspond to nearest-neighbor
spin pairs formed in the directions parallel and perpendicular to the
$S=1/2$ planes, respectively. The variation of the ratio $J'/J$ allows
to study nicely the rich phase diagram of quasi-2D magnetic systems,
including the transition from a~2D Heisenberg AF ($J'/J=0$) to the 3D
Heisenberg AF ($J'/J = 1$). Remarkably, even in presence of relatively
strong interplane interactions, the 2D quantum fluctuations can significantly
affect the magnetic properties of such systems,\cite{sengupta1}
resulting, for instance, in a~peculiar non-monotonic field dependence
of the ordering temperature. Such a behavior has recently been observed
in the quasi-2D square-lattice Heisenberg AF [Cu(C$_4$H$_4$N$_2$)$_2$(HF$_2$)]BF$_4$,\cite{Manson,sengupta1}
and in the spatially anisotropic triangular magnet Cu(tn)Cl$_2$,\cite{betka} where tn is 1,3-diaminopropane (C$_3$H$_{10}$N$_2$).%

Here, we report on magnetic properties of single-crystalline samples
of [Cu(pyz)$_2$(HF$_2$)]PF$_6$, which is regarded as a nearly perfect
realization of a quasi-2D $S=1/2$ Heisenberg AF. Comprehensive
specific-heat, high-field magnetization and electron spin resonance (ESR)
studies allow us to identify a long-range ordered N\'{e}el state with
easy-plane anisotropy and to accurately estimate the spin-Hamiltonian
parameters. We show that in spite of the onset of the 3D ordered state
below $T_N$, the magnetic properties of [Cu(pyz)$_2$(HF$_2$)]PF$_6$ are
still significantly affected by 2D spin correlations.

\section{Experimental details}

\begin{figure}[t]
\includegraphics[width=0.35\textwidth,clip=true]{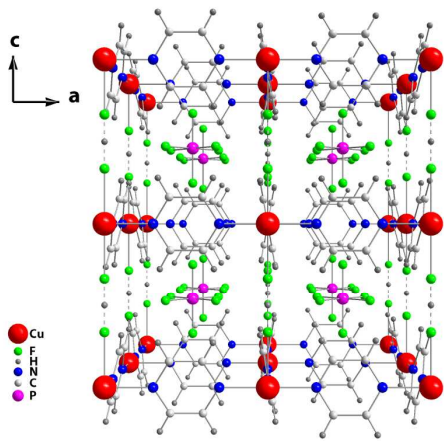}
\caption{\label{fig:structure}(Color online) Structure of [Cu(pyz)$_2$(HF$_2$)]PF$_6$ viewed along the $b$ axis. }
\end{figure}

The measurements have been performed at the Dresden High Magnetic Field
Laboratory ({\it Hochfeld-Magnetlabor Dresden}, HLD). The pulsed-field
magnetization was obtained by integrating the voltage induced in a
compensated pick-up coil system containing a sample. The low-field magnetization below 7~T was measured using a~commercial SQUID magnetometer. The specific-heat measurements were performed using a continuous relaxation-time
technique.\cite{Junod,Lortz}
For the X-band ESR measurements a Bruker ELEXSYS E500 spectrometer
operating at a frequency of 9.4~GHz was used. The ESR experiments
in the frequency range $36 - 250$~GHz were performed at temperatures
down to 1.3~K using a~tunable-frequency home-made ESR spectrometer
(similar to that described in Ref.\ \onlinecite{spectrometer}) equipped
with a 16~T high-homogeneity superconducting magnet. In our
experiments we used single crystals with a typical size of 4x3x1~mm$^3$, synthesized by aqueous reaction of stoichiometric amounts of ammonium
bifluoride, pyrazine, and copper(II) hexafluorophosphate hydrate (using the same procedure as reported in Ref.~\onlinecite{Manson,Manson2})

\section{Results and discussion}
[Cu(pyz)$_2$(HF$_2$)]PF$_6$, where pyz is pyrazine (C$_4$H$_4$N$_2$),
crystallizes in a~tetragonal lattice (space group $P4/nmm$; $Z=1$). The structure consists
of square-lattice planes of Cu$^{2+}$ ions bridged by pyz groups in the
$ab$ plane (Fig.\ \ref{fig:structure}). The planes are connected by HF$_2$ groups along the $c$ axis.
The planes of the pyz groups are nearly orthogonal to the Cu$^{2+}$ planes. The non-coordinating PF$_6$ groups
pack in between the planes, in the center of almost cubic units.~\cite{goddard}

\begin{figure}[t]
\includegraphics[width=0.45\textwidth,clip=true]{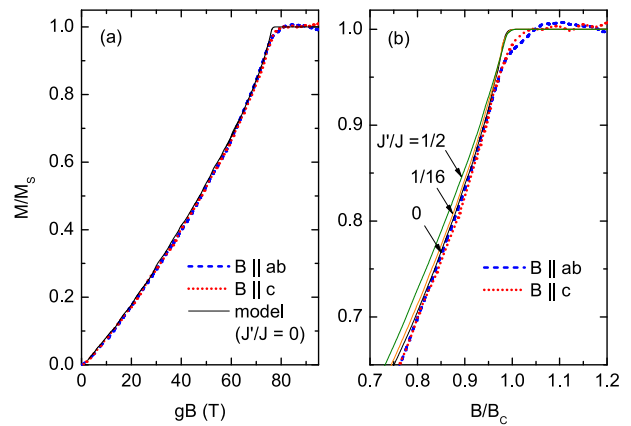}
\caption{\label{fig:magnet}(Color online) (a) High-field magnetization
of [Cu(pyz)$_2$(HF$_2$)]PF$_6$ (dashed and dotted lines correspond to
$B||ab$ and $B||c$, respectively) as a function of $gB$, where $B$ is
the applied magnetic field and the $g$-factors are $g_c=2.28$ and
$g_{ab}=2.05$, as determined by X-band ESR measurements.
The calculated magnetization for an $S=1/2$ 2D
square-lattice Heisenberg AF is shown by the solid line.
(b) The same data in an expanded scale close to saturation with
additional model calculations assuming a finite interplane
interaction.}
\end{figure}

The magnetization of [Cu(pyz)$_2$(HF$_2$)]PF$_6$
was measured at 1.5 K in magnetic fields up to 50 T applied parallel and
perpendicular to the $c$ axis (Fig.\ \ref{fig:magnet}). The fully
spin-polarized state was observed at $B_{c}^{ab} = 37.5$~T and
$B_{c}^c = 33.8$~T for $B||ab$ and $B||c$, respectively. In the scaled
plot of Fig.\ \ref{fig:magnet}, it becomes obvious that the magnetizations
for different magnetic-field orientations coincide nicely. This means
that the magnetization anisotropy is solely determined by the $g$-factor
anisotropy. The numerically calculated magnetization for an $S=1/2$
2D square-lattice Heisenberg AF with an intraplane exchange interaction
$J/k_B=12.8$~K (Refs.\ \onlinecite{goddard, Woodward}) is in very good
agreement with our data (Fig.\ \ref{fig:magnet}). In order to estimate
the ratio $J'/J$, calculations using this model with different interplane
interactions have been made [Fig.\ \ref{fig:magnet}(b)].\cite{goddard}
By comparison with our data it can be concluded that $J'/J$ must be less
than 1/16.

\begin{figure}[t]
\includegraphics[width=0.45\textwidth,clip=true]{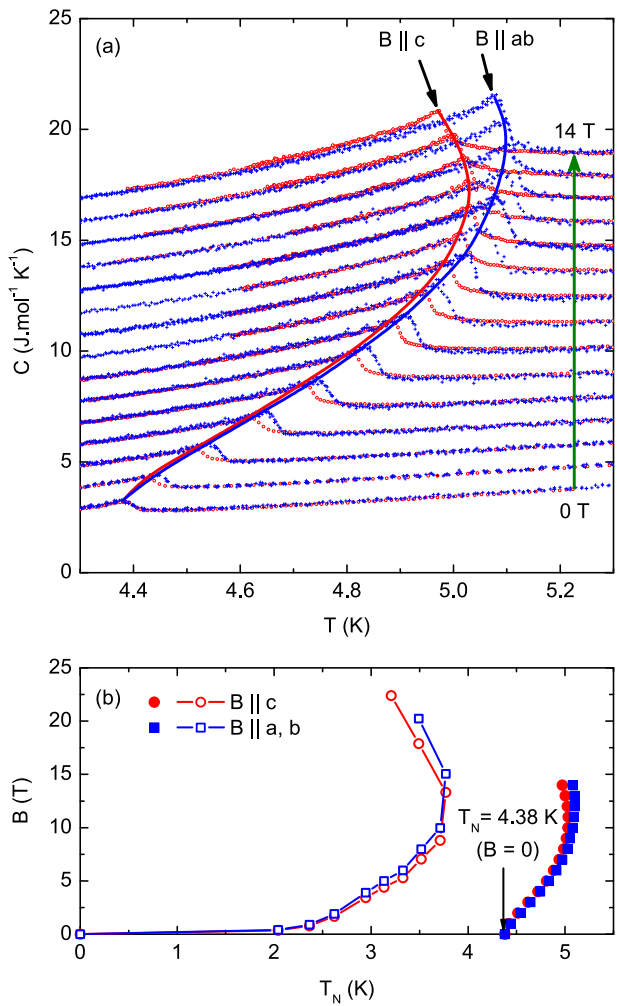}
\caption{\label{fig:phasediag} (Color online) (a) Temperature dependence of
the specific heat of [Cu(pyz)$_2$(HF$_2$)]PF$_6$ for $B||ab$ (blue symbols)
and $B||c$ (red symbols). The experimental data are shifted by
1 Jmol$^{-1}$K$^{-1}$ per Tesla.
(b) The $B-T$ phase diagram of [Cu(pyz)$_2$(HF$_2$)]PF$_6$ obtained
from the specific-heat data (full symbols) shown together with the
calculated phase diagram for the ideal 2D case with $J'=0$
(open symbols).\cite{sengupta1} Lines are guides to the eye.}
\end{figure}

The high-resolution specific-heat measurements allowed to clearly resolve
the 3D AF ordering temperature in zero and applied magnetic fields up to
14~T [Fig.\ \ref{fig:phasediag}(a)]. Using the local maximum of the
specific-heat anomaly, the $B-T$ phase diagram has been extracted.
Figure \ref{fig:phasediag}(b) shows the result together with data
calculated for an ideal 2D Heisenberg AF system ($J'=0$) with
$J/k_B=12.8$~K.\cite{sengupta1} The theory~\cite{cuccoli2,sengupta1}
predicts that an applied magnetic field suppresses fluctuations of the
spin $z$ component. As a result, an effective easy-plane anisotropy is
induced in the system, accompanied by the rise of $T_N$. On the other
hand, for sufficiently strong magnetic field, the spin-canting effect
prevails. Then, the critical temperature decreases and vanishes eventually
at the saturation field, when the fully spin-polarized state is reached.
The measured non-monotonic $B-T$ phase diagram of [Cu(pyz)$_2$(HF$_2$)]PF$_6$
can be nicely explained by modifying above model with a weak interplane
interaction. As a threshold for the observation of a $\lambda$-like
anomaly in the specific heat, $J'/J<0.015$ has been given in
Ref.~\onlinecite{sengupta2}. This ratio is very close to that, $J'/J=0.01$,
estimated from previous experiments on powdered samples.\cite{goddard}

In the range $0.001{\leq}J'/J{\leq}1$ the interplane coupling can be
estimated employing the following formula\cite{yasuda}
\begin{equation}
   T_N = 4\pi\rho_S/\left[2.43-\ln(J'/J)\right],
   \label{eq:ordtemp}
\end{equation}
where $\rho_S=0.183J$ is the spin stiffness for a 2D system. Using the
exchange coupling, obtained from the high-field magnetization data, and
$T_N=4.38$~K we find $J'/J = 0.014$, which is in excellent agreement
with our previous estimate.

\begin{figure}[t]
\includegraphics[width=0.45\textwidth,clip=true]{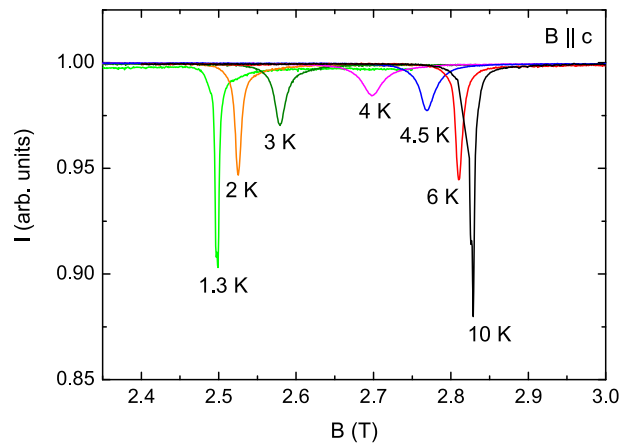}
\caption{\label{fig:spectra} (Color online) ESR signal (transmission) measured at 90~GHz in a temperature range of $1.3-10$~K for $B||c$.}
\end{figure}

ESR is a powerful technique to determine the magnetic structure and the
effective spin-Hamiltonian parameters of  in exchange-coupled spin systems. At room temperature, a single ESR line was observed for all
field orientations with a typical sinusoidal angular dependence of the
$g$-factor expected for a~Cu$^{2+}$ ion in an octahedral surrounding with
an axial symmetry due to the Jahn--Teller effect. The results confirm that
the main exchange paths between the magnetic ions are formed in the $ab$ planes by
$d_{x^{2}-y^{2}}$ orbitals~\cite{AB}, which directly overlap with orbitals of  pyrazine ligands, creating a quasi-2D network of exchange coupled Cu$^{2+}$ ions.

\begin{figure}[t]
\includegraphics[width=0.45\textwidth,clip=true]{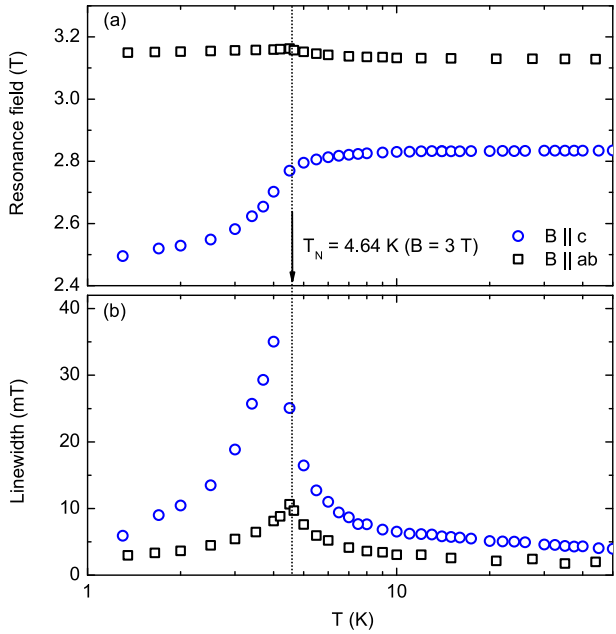}
\caption{\label{fig:eprtdep} (Color online) Temperature dependence of
the resonance field (a) and linewidth (b) obtained from ESR data measured at $\nu=90$~GHz for a magnetic field
aligned parallel (empty squares) and perpendicular (empty circles) to
the $ab$ plane. The transition temperature, $T_N = 4.64$~K, as obtained
from specific-heat measurements at 3~T for $B||ab$ (Fig.\ \ref{fig:phasediag})
is shown by the dotted line.}
\end{figure}

\begin{figure}[t]
\includegraphics[width=0.45\textwidth,clip=true]{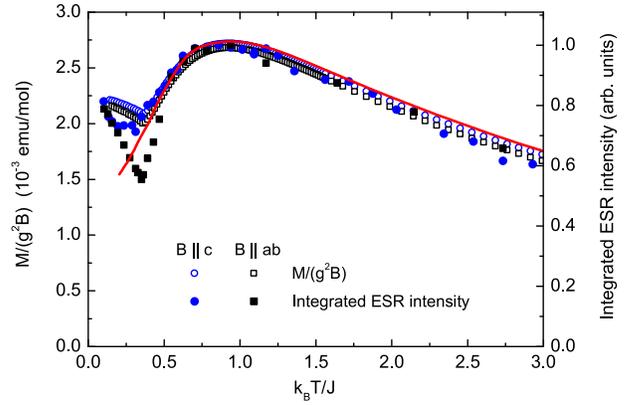}
\caption{\label{fig:susc} (Color online) Temperature dependence of $M/B$ measured at $B=3$~T (open symbols)
and the integrated intensity obtained from ESR data measured at $\nu=90$~GHz (full symbols). The data are
presented together with the calculated results (solid line) for a~2D
square-lattice Heisenberg AF system with $J/k_B = 12.8$~K.\cite{Kim,Woodward}}
\end{figure}

The temperature evolution of the ESR spectra in [Cu(pyz)$_2$(HF$_2$)]PF$_6$
measured at 90 GHz for $B||c$ is shown in Fig.\ \ref{fig:spectra}. The
temperature dependence of the resonance field, the linewidth, and the
integrated intensity of the ESR lines in fields applied parallel and
perpendicular to the $ab$ plane are shown in Fig.\ \ref{fig:eprtdep}(a)-(b) and Fig.\ \ref{fig:susc}.
It is evident that in the vicinity of the ordering temperature
($T_N\approx4.6$~K at 3~T), a region of enhanced short-range spin
correlations is entered and both the resonance field and the linewidth
change dramatically. It is important to mention that the integrated ESR
intensity is proportional to the dynamic spin susceptibility. In Fig.\ \ref{fig:susc}, the temperature dependence of $M/B$ (measured using a~SQUID magnetometer with applied magnetic field $B=3$~T), the integrated ESR intensity, and the magnetic susceptibility calculated for a~2D Heisenberg AF system\cite{Woodward, Kim} are presented. The results of the calculations are in good agreement with our experimental data above $T_N$ for $J/k_B = 12.8$~K, proving the crucial role of 2D spin
correlations in [Cu(pyz)$_2$(HF$_2$)]PF$_6$.

\begin{figure}[b]
\includegraphics[width=0.45\textwidth,clip=true]{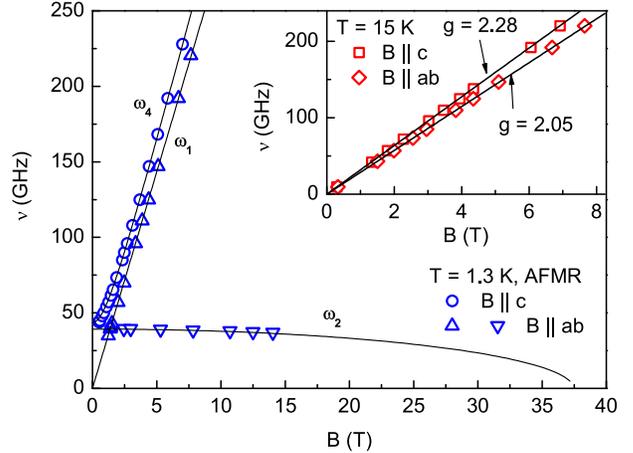}
\caption{\label{fig:epr} (Color online) Frequency-field diagram of the
ESR excitations measured at $T=1.3$~K for a magnetic field aligned parallel
(triangles) and perpendicular (circles) to the $ab$ plane. The signals
$\omega_1$, $\omega_2$, and $\omega_4$ correspond to AF resonance modes
of a 3D ordered AF system with easy-plane anisotropy calculated as
described in the text. The inset shows the ESR modes observed above
the ordering temperature. These modes can be described as transitions
between the energy levels of Cu$^{2+}$ ions by the equation $h\nu=g\mu_{B}B$.}
\end{figure}

The frequency-field diagram of the ESR signals is presented in
Fig.\ \ref{fig:epr}. Well above the critical temperature ($T=15$~K),
we observe only one resonance mode for each orientation, $B||c$ and
$B||ab$, as shown in the inset of Fig.\ \ref{fig:epr}. The frequency-field
dependence of the these modes corresponds to transitions between
the energy levels of the $S=\frac{1}{2}$ Cu$^{2+}$ ions and can be
described by $h\nu=g\mu_{B}B$ with $g_{c} = 2.28$ and $g_{ab} = 2.05$.

The ESR excitation spectrum changes dramatically below $T_N$, as shown for
$T = 1.3$ ~K in Fig.\ \ref{fig:epr}. Two modes, $\omega_1$ and $\omega_2$,
are observed when the magnetic field is applied in the $ab$ plane.
For $B||c$, a nonlinear dependence of the mode $\omega_4$ was observed.
Most importantly, in the ESR excitation spectrum the AF resonance gap
was determined by us directly, $\Delta = 43$~GHz ($\approx2$~K) at $B=0$.
Such a frequency-field diagram is a textbook example for magnetic
excitations in a system with collinear 3D AF long-range order and
easy-plane anisotropy.\cite{Turov} The frequency-field dependence of the
antiferromagnetic resonance modes can be calculated using the mean-field
approximation\cite{AFMR}
\begin{eqnarray}
   \omega_1/{\gamma} = B,
   \label{eq:omegapar}
   \\
   \omega_2/{\gamma} = \sqrt{2B_AB_E-\left({B_A}/{2B_E}\right)B^2},
\end{eqnarray}
for magnetic fields applied parallel to the easy plane and
\begin{eqnarray}
   \omega_3/{\gamma} = 0,
   \\
   \omega_4/{\gamma} = \sqrt{2B_AB_E+\left(1-{B_A}/{2B_E}\right)B^2},
   \label{eq:omegaperp}
\end{eqnarray}
for magnetic fields applied perpendicular to the easy plane, where
$\gamma=g\mu_B/\hbar$ is the gyromagnetic ratio, $B_E$ and $B_A$ are
the exchange and anisotropy field, respectively.
Our results clearly show that below $T_N$ the $ab$ plane is the easy
plane, along which the magnetic moments are aligned at $B=0$. The
frequency-field diagram was analyzed using the Eqs.\ (\ref{eq:omegapar})
-- (\ref{eq:omegaperp}). The best fit was obtained for $B_E=18.72$~T and
$B_A=0.05$~T. The exchange field, $B_E$, is related to the exchange
coupling by the simple relation $B_E=4SJ/\hbar\gamma$ and yields
$J/k_B = 12.9$~K, which is in perfect agreement with the value of
the intraplane exchange coupling obtained from our magnetization and
specific-heat measurements. Noticeably, using ESR  data, the anisotropy constant ($\sim 0.3\%$ of the exchange  interaction) can be estimated, revealing the  validity of  isotropic quasi-2D S=1/2 Heisenberg model applied to  [Cu(pyz)$_2$(HF$_2$)]PF$_6$.

\section{Conclusion}
Our comprehensive specific-heat, high-field magnetization, and ESR
studies revealed the important role of 2D spin correlations on the
magnetic properties of the $S=1/2$ quasi-2D Heisenberg AF
[Cu(pyz)$_2$(HF$_2$)]PF$_6$. The observation of two AF resonance modes below $T_N$ confirms the
presence of a 3D collinear AF long-range ordered state with an energy
gap of $\Delta=43$~GHz in the magnetic excitation spectrum and $A/J\sim 0.003$ for the anisotropy/exchange interaction ratio.  The presence of a~field-induced XY behavior  has unambiguously been identified by the observation of
a~peculiar non-monotonic field dependence of the ordering temperature.

\section*{Acknowledgments}
This work has been supported by Deutsche Forschungsgemeinschaft, EuroMagNET
(EU contract No. 228043), and UChicago Argonne, LLC, Operator of Argonne National
Laboratory ("Argonne"). Argonne, a~U.S. Department of Energy Office of
Science laboratory, is operated under Contract No. DE-AC02-06CH11357. E.\v{C}. is supported also by APVV-VVCE-0058-07
and VEGA 1/0078/09. We would like to thank L.~Zviagina for preparations of pulse-field magnetization measurements.

\end{document}